\documentclass[aps,prb,twocolumn]{revtex4}  
\usepackage{epsfig}
\usepackage{amsmath}
\usepackage{ulem}
\newcommand{\DM}{Dzyaloshinskii-Moriya }
\begin{document}
\title{Quantum phase transition induced by Dzyaloshinskii-Moriya in the kagome antiferromagnet} \author{O.~C\'epas$^1$,
  C. M. Fong$^2$, P.~W.~Leung$^2$, C.~Lhuillier$^1$} \affiliation{$1.$
  Laboratoire de physique th\'eorique de la mati\`ere condens\'ee, UMR7600
  CNRS, Universit\'e Pierre-et-Marie-Curie, Paris 6, 75252 Paris cedex 05,
  France. \\ $2.$ Department of Physics, Hong Kong University of Science and
  Technology, Clear Water Bay, Hong Kong.  }
\begin{abstract}
We argue that the $S=1/2$ kagome antiferromagnet undergoes a quantum phase
transition when the \DM coupling is increased.  For $D<D_c$ the system is in a
moment-free phase and for $D>D_c$ the system develops antiferromagnetic
long-range order. The quantum critical point is found to be $D_c \simeq 0.1J$
using exact diagonalizations and finite-size scaling. This suggests that the
kagome compound ZnCu$_3$(OH)$_6$Cl$_3$ may be in a quantum critical region
controlled by this fixed point.
\end{abstract}

\pacs{PACS numbers:} \maketitle
\date{\today}

In the search for materials realising a spin-liquid ground state, one has to
face the presence of small anisotropic interactions of spin-orbit origin. Such
interactions that break the full rotation symmetry of the Heisenberg model,
reduce the quantum fluctuations and may tend to induce magnetic phases at low
temperatures. The recently discovered spin $1/2$ copper oxide
ZnCu$_3$(OH)$_6$Cl$_3$\cite{Shores} that has the geometry of a kagome lattice
may be a good candidate for a spin-liquid.\cite{Today} Experimentally no
apparent freezing of the magnetic moments has been found down to very low
temperatures,\cite{Mendels,Ofer,Helton} despite strong Heisenberg
interactions. Exact diagonalizations of the Heisenberg model predict indeed a
non-magnetic state with no magnetic moment.\cite{Leung,Lecheminant} 
 However, smaller interactions of spin-orbit origin are
certainly present. In particular those of \DM symmetry\cite{DM} are expected
when the magnetic bonds have no inversion center, which is the case of
ZnCu$_3$(OH)$_6$Cl$_3$.\cite{Shores} An immediate question is to what extent
they affect the non-magnetic phase.

\DM interactions have been first invoked\cite{Rigol} in the context of
ZnCu$_3$(OH)$_6$Cl$_3$ to explain the enhancement of the spin susceptibility
at low temperatures.\cite{Helton} NMR measurements of the local susceptibility
have provided a different interpretation in terms of the presence of defects
in the structure,\cite{Olariu} a result consistent with a direct fit of the
susceptibility,\cite{Misguich} and corroborated by theoretical calculations.\cite{Rozenberg} A more direct evidence of anisotropy has been
obtained by paramagnetic resonance.\cite{Zorko} A \DM coupling of order
$0.08J$ was needed to explain the linewidth,\cite{Zorko} which is typical of
cuprates.  It is therefore a relatively small correction but may be of crucial
importance in highly frustrated systems. In fact it is known that an infinitesimally small \DM interaction in the \textit{classical} kagome favors long-range
N\'eel order with a $\mathbf{Q}=0 $ propagation vector and 120$^o$ orientation
of the spins.\cite{Elhajal} Spin-wave corrections renormalize down the
magnetic moment but do not suppress it for $D=0.1J$.\cite{Ballou} So it was
unclear how this N\'eel phase could be reconciled with experimental
observations.

In this letter, we show that the proper inclusion of quantum fluctuations
leads to a phase transition from a N\'eel state to a moment-free phase at a
quantum critical point that we estimate to be $D_c \simeq 0.1J$.  This
therefore resolves the contradiction for ZnCu$_3$(OH)$_6$Cl$_3$ and,
furthermore, suggests that its magnetic properties may be strongly influenced
by the proximity of the quantum critical point.  In particular power-law
scalings were observed in the dynamical susceptibility\cite{Helton} and NMR
relaxation times,\cite{Imai,Olariu} and have been interpreted so far in terms
of \textit{critical} spin liquid states for the kagome.\cite{Ran,Ryu} The
present results indicate that the power-laws of ZnCu$_3$(OH)$_6$Cl$_3$
may well originate in the present fixed point.  In order to identify the phases in an
unbiased way and locate $D_c$, we have performed exact diagonalizations of
small clusters.  At $D=0.1J$, there is a clear emerging low-energy
\textit{tower of states} that collapses onto the ground state like $1/N$, thus
signaling a broken-symmetry phase. We have calculated the N\'eel
order-parameters as function of $D/J$ and have found that they vanish for
$D\lesssim 0.1J$, thus leading to a moment-free phase and $D_c \simeq 0.1J$.
\begin{figure}[htbp]
\centerline{
 \psfig{file=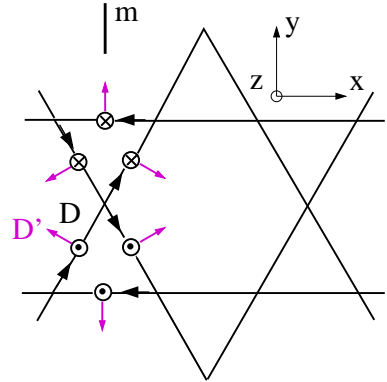,width=5cm,angle=0}}
\caption{(color online). The kagome lattice with the allowed Dzyaloshinskii-Moriya
  Interactions. The orientation of
  the bonds specify the order of the operators in $ \textbf{\mbox{S}}_i \times \textbf{\mbox{S}}_j $. m is a mirror plane.}
\label{Lattice}
\end{figure}

The model we are considering is based on the symmetries of the two-dimensional
kagome lattice (Fig.~\ref{Lattice}),\cite{Elhajal}
\begin{equation}
H = \sum_{nn} \left[ J \textbf{\mbox{S}}_i . \textbf{\mbox{S}}_j  + \textbf{\mbox{D}}
_{ij}. ( \textbf{\mbox{S}}_i \times \textbf{\mbox{S}}_j) \right]
\label{ham}
\end{equation}
where $nn$ stands for nearest neighbors and $\mathbf{S}_i$ is a
$S=1/2$ quantum spin on site $i$. Thanks to the lack of inversion
symmetry at the middle of each bond, interactions of \DM symmetry are
allowed between nearest neighbors. According to Moriya's
rules,\cite{DM} there are components of $\mathbf{D}_{ij}$
perpendicular to the planes, of strength $D$, staggered from up
triangles to down triangles; and in-plane components that point
towards the center of each triangle of strength $D^{\prime}$ (see
Fig.~\ref{Lattice}).\cite{Elhajal} There is \textit{a priori} no simple
relation between $D$ and $D^{\prime}$. It is true that if the
Cu-(OH)-Cu plane was a mirror plane of the crystal structure, then the
$\mathbf{D}$ vector would be perpendicular to it. In fact the proton
of (OH) breaks that symmetry,\cite{Shores} and since the perturbation
is expected to be strong it is difficult to relate $D$ and $D'$. For
ZnCu$_3$(OH)$_6$Cl$_3$, $J$ has been estimated from the susceptibility
to be about 170-190 K.\cite{Rigol,Misguich} Interplane couplings are
thought to be smaller because copper ions are far away in the $c$
direction and separated by sheets of zinc and will be neglected in the
following. According to electron spin resonance measurements,
$D=15$ K ($D \sim 0.08J$) and $D^{\prime}=2$ K ($D^{\prime} \sim
0.01J$).\cite{Zorko}

We now map the model (\ref{ham}) onto a simpler model that restores a U(1)
symmetry up to terms of order $D'^2/J$. The in-plane components vectors sum up
to zero when going around a triangle and are therefore \textit{reducible} to a
$D'^2/J$ term (for small $D^{\prime}$)\cite{Cheng} by appropriate rotations of
the spin operators.\cite{Aharony} We shall neglect these second-order terms not only because
they are smaller but also because there are other (symmetric) exchange anisotropies at the
same order that we have not included.  Therefore in the rotated frame, the
model has only the original $D$ component along $z$ with the same strength and
has a U(1) rotation symmetry about this axis: it is this symmetry that, as
we shall show, is going to be spontaneously broken. In the following we shall
present the results of the numerical diagonalization of (\ref{ham}) for
systems of size $N=21,24,27,30,36$ (for the $N=24$ to $27$, we have
considered 2 different cluster shapes). The dimension of the largest Hilbert space is 
$\sim 7\times 10^8$. We have started with a \DM coupling strength fixed to $D=0.1J$ (and $J=1$). 
\begin{figure}[htbp]
\centerline{
 \psfig{file=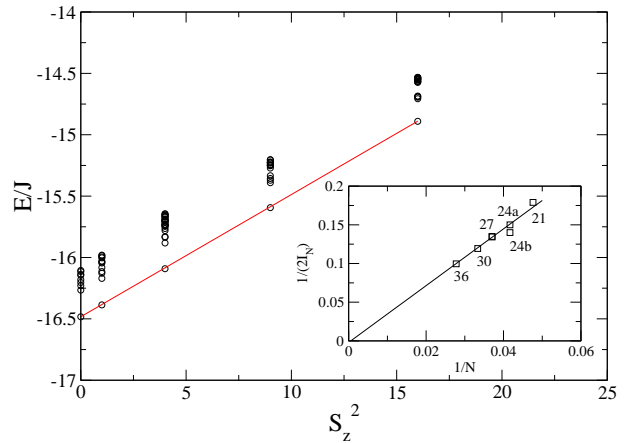,width=7cm,angle=-90}}
\caption{(color online). Low-energy levels of the Heisenberg kagome lattice with \DM
  interactions for $N=36$. The inset gives the slope of the lowest energy states
  versus $S_z^2$, denoted by $1/(2I_N)$, as a function of $1/N$. ($D=0.1J$).}
\label{lowenergylevels}
\end{figure}

\textit{Symmetry breaking in the thermodynamic limit}. We have calculated
several low-energy levels of (\ref{ham}) in each sector of the total
magnetization, denoted by $S_z$, for different cluster sizes. For $N=36$, the
energy levels are shown in Fig.~\ref{lowenergylevels}. The spectrum is
qualitatively different from the exact spectra obtained at $D=0$ (see
refs.~[\onlinecite{Lecheminant}]). Here we clearly see a band of low energy
levels (red line in Fig.~\ref{lowenergylevels}) well separated from higher
energy states. It forms a so-called \textit{tower of states} which energy is
very well described by a quadratic term $S_z^2/2I_N$
(Fig.~\ref{lowenergylevels}). The slope $1/(2I_N)$ is fitted for all available
$N$ and shown in the inset of Fig.~\ref{lowenergylevels}. It clearly
extrapolates to zero in the thermodynamic limit like $1/N$. 
In this limit one can then form a superposition of these 
eigenstates with different $S_z$, thus forming a 
macroscopic state with a preferred in plane direction. This therefore 
shows that the system breaks the rotation symmetry in the thermodynamic 
limit. In this standpoint, the degeneracy of the tower of states is then 
a natural consequence of the invariance of the Hamiltonian in rotations 
about the internal  axis defined by $\mathbf{D}$:  a given macrosopic 
state pointing in the $\textbf{u}$ direction in the transverse plane is 
degenerate with all its transforms in any rotation about $\mathbf{D}$.
In addition, the wave-vector of the lowest energy state in each $S_z$ sector is
$\mathbf{Q}=0$ for all clusters, so that we can safely conclude that the
system will not break translation invariance in the thermodynamic limit.
Furthermore, because of the continuous broken-symmetry, we expect a
long wavelength Goldstone mode with energy varying like $k \propto
N^{-1/2}$ in two dimensions. Unfortunately, the first allowed $k \neq
0$ wave-vectors are not small enough on these clusters to observe the
individual long wavelength states directly in the spectra. In fact,
indirect confirmation of a Goldstone mode will be given below in the
scalings of the energy of the groundstate and correlations.
From the spectra, we can now
extract the uniform susceptibility at T=0 by using the
expression of the energy, $S_z^2/2I_N - HS_z$. The susceptibility of the
ground state $I_N$ is proportional to $N$ (as shown in
Fig.~\ref{lowenergylevels}) and the susceptibility per site $\chi=I_N/N$ is
found to be $\chi=0.144 \pm 0.002$ for $D/J=0.1$ in the thermodynamic limit.

\begin{figure}[htbp]
\centerline{ \psfig{file=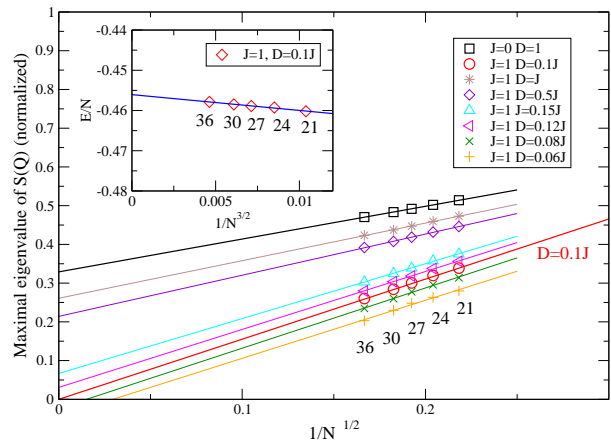,width=7cm,angle=-90}}
\caption{(color online). Finite-size extrapolation of the largest eigenvalue of
  $S_{ab}(\mathbf{Q}=0)$, showing \textit{(i)} a scaling in $N^{-1/2}$;
  \textit{(ii)} the decrease of the sublattice moment from $m_{AF}^2=0.326$ ($D=1$,
  $J=0$) to zero ($D=0.1J$) and negative values ($D=0.06;0.08J$).  Inset: finite-size
  scaling of the ground state energy in $N^{-3/2}$ for $D=0.1J$. }
\label{fsc}
\end{figure}

\begin{figure}[htbp]
\centerline{
 \psfig{file=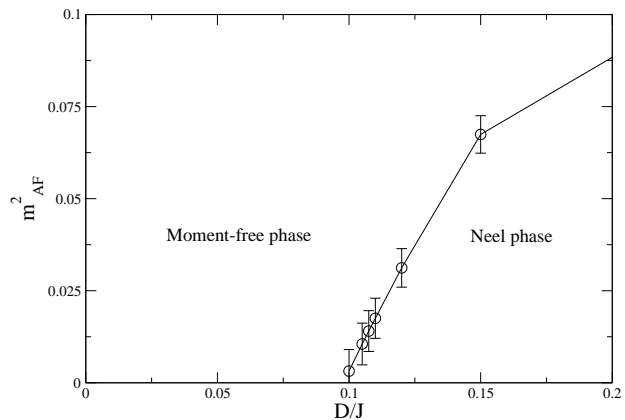,width=7cm,angle=-90}}
\caption{ The N\'eel order-parameter as a function of $D/J$,
  obtained from extrapolations to infinite size. The system has no magnetic
  moment for $D<D_c \simeq 0.1J$ and has N\'eel order for $D>D_c$. }
\label{phasediagram}
\end{figure}

\textit{N\'eel order-parameters}. To test for the possibility to have a
sublattice magnetization (or N\'eel order) in the thermodynamic limit, we have
calculated the spin-spin correlations in the (finite-size) ground state. We
distinguish between correlations in a plane perpendicular to $\mathbf{D}$, $
\langle 0 | S_{ia}^x S_{j b}^x + S_{ia}^y S_{j b}^y | 0 \rangle $, and the correlations
along $\mathbf{D}$, $\langle 0|S_{i a}^{z} S_{j b}^{z} | 0\rangle$. As
expected from the easy-plane character of the \DM interaction, the latter
remains much smaller and especially at large separation.
To study the in-plane ordering, we define the Fourier transform of 
 the spin-spin correlations:
\begin{equation}
S_{ab}(\mathbf{Q}) = \frac{24}{N^2} \sum_{ij}  e^{i \mathbf{Q} \cdot
  (\mathbf{R}_i-\mathbf{R}_j)}  \langle 0 |
S_{ia}^x S_{j b}^x  | 0 \rangle
\label{op}
\end{equation}
This is a $3\times 3$ matrix that turns out to be peaked at
$\mathbf{Q}=0$. The largest eigenvalue at $\mathbf{Q}=0$ corresponds simply to
the 120$^o$ in-plane orientation of the magnetic moments within the
unit-cell. The prefactor 24 is chosen so that the largest eigenvalue is 1 in
the perfect N\'eel state. For a quantum magnet with a finite N\'eel
order-parameter, the eigenvalue should extrapolate to a finite value smaller
than 1 (because of quantum fluctuations), with a well-defined finite-size
scaling.  The latter can be predicted from the existence of a low energy spin
wave with wavevector varying like $N^{-1/2}$ in two dimensions, and is of the
form $N^{-1/2}$ for the correlations and $N^{-3/2}$ for the energy per
site.\cite{fss} Fig.~\ref{fsc} shows the largest eigenvalue versus $N^{-1/2}$
for different $D/J$. First we see that the scaling is obeyed both for the
correlations (Fig.~\ref{fsc}) and for the groundstate energy per site (inset
of Fig.~\ref{fsc}).  Second, the intercept for infinite system-size gives the
order-parameter, noted $m_{AF}^2$. Starting from large $D/J$ (squares) and
reducing $D/J$ we see that the order-parameter decreases from $m_{AF}^2 \sim
0.326$ to zero at $D_c \simeq 0.1J$ (circles). For $D<D_c$, it is no longer
possible to find a finite order-parameter: the negative extrapolated values
reflect the breakdown of the scaling at some length-scale and short-range
correlations develop instead.  In Fig.~\ref{phasediagram}, we summarize the
behavior of the N\'eel order-parameter $m_{AF}$, as a function of $D/J$.  The
order-parameter decreases continuously to zero, so that the transition is
compatible with a second-order phase transition.  It seems difficult though to
extract an accurate value for the critical exponent, given the error bars resulting from
finite-size scaling. The present data are compatible with the mean-field
behavior, $m_{AF}^2 \propto (D-D_c)$ but the exponent could also be
smaller. In any case, $D_c \simeq 0.1J$ clearly appears as the critical point
separating the N\'eel phase ($Q=0$, 120$^o$ in-plane orientation of the
spins)\cite{note} from a phase with no static moment. 
It is a more accurate estimate than that of
spin-waves (that is three times smaller),\cite{Ballou} because here all
quantum fluctuations have been taken into account.


In order to have a simple understanding for the occurrence of a N\'eel phase in
the phase diagram, we discuss the limit of large $D/J$, for which simple
physical arguments can be used.  For this we start with a trimerized version
of the kagome lattice.\cite{trimerized} On a triangle, three Heisenberg
quantum spins 1/2 form a quartet and two degenerate low-energy doublets. The
\DM interaction lifts the degeneracy and selects the doublet with the
(vector)-chirality opposite to $\mathbf{D}$.  We now assume that $D$ is large
and ignore the higher doublet (and quartet). We define a pseudo-spin
$\sigma_i=1/2$ for the lowest doublet on each triangle. The inter-triangle \DM
interaction (that is now supposed to be large) gives an effective interaction
for the pseudo-spins that we can write
\begin{equation}
H' =  -  \frac{2}{9} |D| \sum_{<i,j>} \sqrt{3} (\sigma_i^x \sigma_j^x +
\sigma_i^y \sigma_j^y) + \mathbf{\hat{e}}_{ij}\cdot ( \mathbf{\sigma}_i \times \mathbf{\sigma}_j)  
\label{effective}
\end{equation}
where $i,j$ are now sites on a triangular lattice. $\mathbf{\hat{e}_{ij}}$ is
a unit-vector along $+z$ when turning anticlockwise around an up triangle of
the triangular lattice. (\ref{effective}) is essentially a
\textit{ferromagnetic} in-plane interaction with a sizeable effective \DM
interaction. The latter gives magnetic frustration and the problem of solving
(\ref{effective}) is in principle as complicated as the original problem. For
three triangles, the ground state of (\ref{effective}) is the $\mathbf{Q}=0$
doublet state (of pseudo-spins), made of $(|\uparrow \uparrow \downarrow
\rangle + |\uparrow \downarrow \uparrow \rangle +|\downarrow \uparrow \uparrow
\rangle)/\sqrt{3}$ and its time reversal counterpart. Similarly, the classical
ground state of (\ref{effective}) is the $\mathbf{Q}=0$ in-plane ferromagnetic
state with \textit{e.g.} $\langle \sigma_i^x\rangle= 1/2 $ on all
triangles. In terms of the original spins, this corresponds to long-range
N\'eel order in $\langle \mathbf{S}_{ia} \rangle$, of the $\mathbf{Q}=0$,
120$^o$ form.  The reduction of the in-plane moment can also be estimated. For
the three-triangle ferromagnetic doublet given above, one can form a
superposition of the two states that gives a moment along $x$, for instance.
The moment is then found to be a fraction $m_{AF}=4/9 \sim 0.44$ of the full
moment. This gives a simple explanation for the $\mathbf{Q}=0$ N\'eel phase
found by exact diagonalization in the limit of large $D/J$. Since we know from
previous works that the $D=0$ phase is non-magnetic,\cite{Leung,Lecheminant}
we naturally expect at least one quantum critical point between the two
phases.

We conclude that the $S=1/2$ kagome antiferromagnet has a quantum critical
point at $D_c \simeq 0.1J$ separating a moment-free phase ($D<D_c$) from a
N\'eel phase with a sublattice moment ($D>D_c$) (or a weak net moment if
$D^{\prime} \neq 0$, see [\onlinecite{note}]). This is clearly compatible with
the absence of a static moment in
ZnCu$_3$(OH)$_6$Cl$_3$,\cite{Mendels,Ofer,Helton} because the coupling
extracted from ESR, $D=0.08J$,\cite{Zorko} is smaller than the critical
coupling.\cite{notesus} Furthermore this estimation places this compound very
close to the quantum critical point.  This raises the issue of the origin of
the power-law behaviors observed experimentally when the temperature is
decreased.\cite{Helton,Olariu,Imai} They have been interpreted so far in terms
of \textit{critical} spin liquids,\cite{Ryu,Ran} or free $S=1/2$
impurities.\cite{Chitra} Here we suggest a different intrinsic interpretation
in terms of the proximity with the present critical point.  The prediction of
the critical behavior and low-temperature scalings is not an easy task,
however. In the view of the present uncertainties about the $D=0$ phase and
the other possibilities of valence bond crystal states\cite{Huse} or
intermediate phases such as a spin nematic state, for instance, it is indeed
difficult to ascertain what the effective low-energy quantum field theory
is. It is interesting to note though that the same ordered phase was found as
an instability of the field theory describing the algebraic spin
liquid,\cite{Hermele} but here we have given a finite critical value for
$D_c$.  At finite temperatures such a critical point will open a
quantum critical region that may have consequences on the magnetic properties
of ZnCu$_3$(OH)$_6$Cl$_3$. It is also of course of particular interest if the
compound could be driven across the transition by applying an external
pressure or magnetic field, for instance.

We would like to thank F.~Bert, P. Mendels and A.~Zorko for sharing with us
unpublished experimental data, and B.~Canals, C.~Lacroix, A. Laeuchli and F.~Mila for discussions. O.C. would like to thank the ILL and the Institut N\'eel for hospitality.
C.~M.~F. and P.~W.~L. are supported by the Hong Kong RGC Grant No. 601207.


\begin{thebibliography}{99}
\bibitem{Shores} M. P. Shores, E. A. Nytko, B. M. Bartlett, D. G. Nocera, J. Am. Chem Soc. \textbf{127}, 13462 (2005).
\bibitem{Today} B. Levi, Physics Today \textbf{60}, 16 (2007).
\bibitem{Mendels} P. Mendels, F. Bert, M. A. de Vries, A. Olariu, A. Harrison, F. Duc, J. C. Trombe, J. S. Lord, A. Amato, and C. Baines, Phys. Rev. Lett. \textbf{98}, 077204 (2007).
\bibitem{Ofer} O. Ofer, A. Keren, E. A. Nytko, M. P. Shores, B. M. Bartlett,
   D. G. Nocera, C. Baines, A. Amato,  condmat/0610540 (unpublished).
\bibitem{Helton}
J. S. Helton, K. Matan, M. P. Shores, E. A. Nytko, B. M. Bartlett, Y. Yoshida, Y. Takano, A. Suslov, Y. Qiu, J.-H. Chung, D. G. Nocera, and Y. S. Lee, Phys. Rev. Lett. \textbf{98}, 107204 (2007).
\bibitem{Leung} P. W. Leung and V. Elser, Phys. Rev. B \textbf{47},
5459 (1993).
\bibitem{Lecheminant} P. Lecheminant, B. Bernu, C. Lhuillier, L. Pierre, and P. Sindzingre, Phys. Rev. B \textbf{56}, 2521
(1997);  C. Waldtmann, H.-U. Everts, B. Bernu, C. Lhuillier, P. Sindzingre, P. Lecheminant, and L. Pierre, Eur. Phys. J. B \textbf{2}, 501 (1998).
\bibitem{DM} I. Dzyaloshinskii, J. Phys. Chem. Solids \textbf{4}, 241 (1958);
  T.~Moriya, Phys. Rev. \textbf{120}, 91 (1960).
\bibitem{Rigol} M. Rigol and R. R. P. Singh, Phys. Rev. Lett. \textbf{98},
  207204 (2007); M. Rigol and R. R. P. Singh, Phys. Rev. B \textbf{76}, 184403 (2007).
\bibitem{Olariu} A. Olariu, P. Mendels, F. Bert, F. Duc, J. C. Trombe, M. A. de Vries, and A. Harrison, Phys. Rev. Lett. \textbf{100},
  087202 (2008).
\bibitem{Misguich} G. Misguich and P. Sindzingre, Eur. Phys. J. B \textbf{59}, 305 (2007).
\bibitem{Rozenberg} M.~J.~Rozenberg and R. Chitra, cond-mat/0805.3483 (unpublished).
\bibitem{Zorko} A. Zorko, S. Nellutla, J. van Tol, L. C. Brunel, F. Bert,
  F. Duc, J. C. Trombe, M. A. de Vries, A. Harrison, and P. Mendels, cond-mat/0804.3107 (unpublished).
\bibitem{Elhajal} M. Elhajal, B. Canals, and C. Lacroix, Phys. Rev. B \textbf{66}, 014422 (2002).
\bibitem{Ballou}  M. Elhajal, Ph.D. thesis, University Joseph Fourier,
  Grenoble, 2002 (unpublished); R. Ballou, B. Canals, M. Elhajal, C. Lacroix, and A. S. Wills, J. Magn. Magn. Mat. \textbf{262},
  465 (2003).
\bibitem{Imai}  T. Imai, E. A. Nytko, B. M. Bartlett, M. P. Shores, and
  D. G. Nocera, Phys. Rev. Lett. \textbf{100}, 077203 (2008).
\bibitem{Ran} Y.~Ran, M.~Hermele, P.~A.~Lee, and X.-G.~Wen,
  Phys. Rev. Lett. \textbf{98}, 117205 (2007).
\bibitem{Ryu}  S. Ryu, O. I. Motrunich, J. Alicea, and M. P. A. Fisher,
  Phys. Rev. B \textbf{75}, 1884406 (2007).
\bibitem{Cheng}  Y.~F.~Cheng, O. C\'epas, P. W. Leung, and T. Ziman, Phys. Rev. B \textbf{75},
  144422 (2007).
\bibitem{Aharony} L. Shekhtman, O. Entin-Wohlman, and A. Aharony, Phys. Rev. Lett. \textbf{69}, 836 (1992).
\bibitem{fss}  H. Neuberger and T. Ziman, Phys. Rev. B \textbf{39}, 2608
  (1989); D. S. Fisher, Phys. Rev. B \textbf{39}, 11783 (1989).
\bibitem{note} When $D^{\prime} \neq 0$, we have to
  transform the operators back onto the original frame: the
  120$^o$ order transforms onto a weak ferromagnet. In fact the classical
  energy difference between the weak ferromagnet and the coplanar 120$^o$
  state is only of order $D^{\prime 2}$, and one must include second-order
  exchange anisotropies to actually predict the form of order.\cite{Aharony} 
\bibitem{trimerized} V. Subrahmanyam, Phys. Rev. B \textbf{52}, 1133 (1995);
  F. Mila, Phys. Rev. Lett. \textbf{81}, 2356 (1998).  
\bibitem{notesus}  Note also that our estimate of the zero-temperature susceptibility 
at $D=D_c$, $\chi =0.144$ is close to the intrinsic susceptibility measured by NMR, $\chi=0.13$.\cite{Olariu} 
\bibitem{Chitra} R. Chitra, and M. J. Rozenberg, Phys. Rev. B \textbf{77}, 052407 (2008).
\bibitem{Huse} J. B. Marston and C. Zeng, J. Appl. Phys. \textbf{69}, 5962
  (1991); C. Zeng and V. Elser,  Phys. Rev. B \textbf{51}, 8318 (1995);
  A. V. Syromyatnikov and S. V. Maleyev, Phys. Rev. B \textbf{66}, 132408
  (2002); R. Budnik and A. Auerbach, Phys. Rev. Lett. \textbf{93}, 187205
  (2004); P. Nikoli\'c and T. Senthil, Phys. Rev. B \textbf{68}, 214415
  (2003); R. R. P. Singh and D. A. Huse, Phys. Rev. B \textbf{76}, 180407(R)
  (2007).
\bibitem{Hermele} M.~Hermele, Y.~Ran, P.~A.~Lee, and X.-G.~Wen, Phys. Rev. B \textbf{77},
  224413 (2008).

\end{thebibliography}
\end{document}